\documentclass[bibyear]{aa}
\usepackage{graphicx}
\usepackage{txfonts}
\authorrunning{S.~Basso, B.~Salmaso, D.~Spiga et al.}
\titlerunning{First light of BEaTriX, the new testing facility}

\begin{document}

   \title{First light of BEaTriX, the new testing facility for the modular X-ray optics of the ATHENA mission}

   \author{S.~Basso\inst{1}, B.~Salmaso\inst{1}, D.~Spiga\inst{1}, M.~Ghigo\inst{1}, G.~Vecchi\inst{1}, G.~Sironi\inst{1}, V.~Cotroneo\inst{1}, P.~Conconi\inst{1}, E.~Redaelli\inst{1}, A.~Bianco\inst{1}, G.~Pareschi\inst{1}, G.~Tagliaferri\inst{1}, D.~Sisana\inst{2}, C.~Pelliciari\inst{3}, M.~Fiorini\inst{4}, S.~Incorvaia\inst{4}, M.~Uslenghi\inst{4}, L.~Paoletti\inst{5}, C.~Ferrari\inst{6, 1}, R.~Lolli\inst{6}, A.~Zappettini\inst{6}, M.~Sanchez~del~Rio\inst{7}, G.~Parodi\inst{8}, V.~Burwitz\inst{9}, S.~Rukdee\inst{9}, G.~Hartner\inst{9}, T.~M\"uller\inst{9}, T.~Schmidt\inst{9}, A.~Langmeier\inst{9}, D.~Della~Monica~Ferreira\inst{10}, S.~Massahi\inst{10}, N.C.~Gellert\inst{10}, F.~Christensen\inst{10}, M.~Bavdaz\inst{11}, I.~Ferreira\inst{11}, M.~Collon\inst{12}, G.~Vacanti\inst{12}, N.M.~Barri\`ere\inst{12}}

   \institute{Istituto Nazionale di AstroFisica (INAF), Osservatorio Astronomico di Brera (OAB), Via E. Bianchi 46, 23807 Merate (Italy)
                  \and Politecnico Milano Bovisa, Via La Masa 34, 20156 Milano (Italy)
                  \and IIS Bachelet, Via Stignani 63/65, 20081 Abbiategrasso, Milano (Italy)
                  \and INAF-IASF Milano, Via A. Corti 12, 40133 Milano (Italy)
                  \and INAF Astronomical Observatory Padova, Vicolo Osservatorio 5, 35122 Padova (Italy)
                  \and IMEM-CNR, Parco Area delle Scienze 37/A, 43124 Parma (Italy)
                  \and European Synchrotron Radiation Facility, B.P. 220, 38043 Grenoble (France)
                  \and BCV Progetti, Via S. Orsola 1, 20123 Milano (Italy)
                  \and Max-Planck-Institut f\"ur extraterrestrische Physik, Giessenbachstr, 85748 Garching (Germany)
                  \and DTU-space, Juliane Maries Vej 30, DK-2100 Copenhagen (Denmark)
                  \and ESTEC, European Space Agency, Keplerlaan 1, 2201 AZ Noordwijk (The Netherlands)
                  \and cosine, Oosteinde 36, 2361 HE Warmond, (The Netherlands)}

   \date{Received 16 May 2022 / Accepted 29 June 2022}
 
  \abstract
   {}
   {The Beam Expander Testing X-ray facility (BEaTriX) is a unique X-ray apparatus now operated at the Istituto Nazionale di Astrofisica (INAF), Osservatorio Astronomico di Brera (OAB), in Merate, Italy. It has been specifically designed to measure the point spread function (PSF) and the effective area (EA) of the X-ray mirror modules (MMs) of the Advanced Telescope for High-ENergy Astrophysics (ATHENA), based on silicon pore optics (SPO) technology, for verification before integration into the mirror assembly. To this end, BEaTriX generates a broad, uniform, monochromatic, and collimated X-ray beam at 4.51~keV. The beam collimation is better than a few arcseconds, ensuring reliable tests of the ATHENA MMs, in their focus at a 12~m distance.}
   {In BEaTriX, a micro-focus X-ray source with a titanium anode is placed in the focus of a paraboloidal mirror, which generates a parallel beam. A crystal monochromator selects the 4.51 keV line, which is expanded to the final size by a crystal asymmetrically cut with respect to the crystalline planes. An in-house-built Hartmann plate was used to characterize the X-ray beam divergence, observing the deviation of X-ray beams from the nominal positions, on a 12-meter-distant CCD camera. After characterization, the BEaTriX beam has the nominal dimensions of 170~mm $\times$ 60~mm, with a vertical divergence of 1.65~arcsec and a horizontal divergence varying between 2.7 and 3.45~arcsec, depending on the monochromator setting: either high collimation or high intensity. The flux per area unit varies from 10 to 50 photons/s/cm$^2$ from one configuration to the other.}
   {The BEaTriX beam performance was tested using an SPO MM, whose entrance pupil was fully illuminated by the expanded beam, and its focus was directly imaged onto the camera. The first light test returned a PSF and an EA in full agreement with expectations. As of today, the 4.51~keV beamline of BEaTriX is operational and can characterize modular X-ray optics, measuring their PSF and EA with a typical exposure of 30~minutes. Another beamline at 1.49~keV is under development and will be integrated into the current equipment. We expect BEaTriX to be a crucial facility for the functional test of modular X-ray optics, such as the SPO MMs for ATHENA.}
   {}

   \maketitle
   
\section{Introduction}
\label{sect:intro}
Testing optics for an X-ray telescope requires an at-wavelength illumination that mimics an astronomical X-ray source, that is, a parallel, broad, and uniform X-ray beam. Standard laboratory X-ray sources are naturally diverging and therefore have to be placed at several hundreds of meters from the optic being tested to obtain a sufficiently low-divergence, full illumination of the optics (such as at PANTER, the X-ray test facility of the Max-Planck Institute near Munich, Germany; Burwitz et al.~\cite{Burwitz2019}). Since the soft X-ray absorption in air is very high, a very long high-vacuum system has to be built to enable X-ray propagation from the source to the experimental chamber. Moreover, large optics usually require large volumes in high vacuum, which entail a long pumping time prior to any X-ray characterization. Synchrotron light (such as at the BESSY synchrotron facility, in Berlin, Germany) provides the required level of collimation (Krumrey et al.~\cite{Krumrey2016}); however, the beam is usually much narrower than the aperture of the module, and so it has to be scanned through the module aperture to reconstruct the full focal spot.

As an alternative solution, the Beam Expander Testing X-ray facility (BEaTriX), was designed as a compact (9~m $\times$ 18~m) X-ray apparatus with fast high-vacuum pumping, a small experimental chamber, and a unique optical setup that produces a 170~mm $\times$ 60~mm wide, parallel and monochromatic X-ray beam (Spiga et al. \cite{Spiga2012, Spiga2014, Spiga2016, Spiga2019}; Pelliciari et al. \cite{Pelliciari2015}; Salmaso et al. \cite{Salmaso2018, Salmaso2019, Salmaso2021}). BEaTriX is suitable to fully illuminate the aperture of X-ray modular optics, and thus test their focusing performances. 

\begin{figure}
   \centering
   \includegraphics[width= 0.48\textwidth]{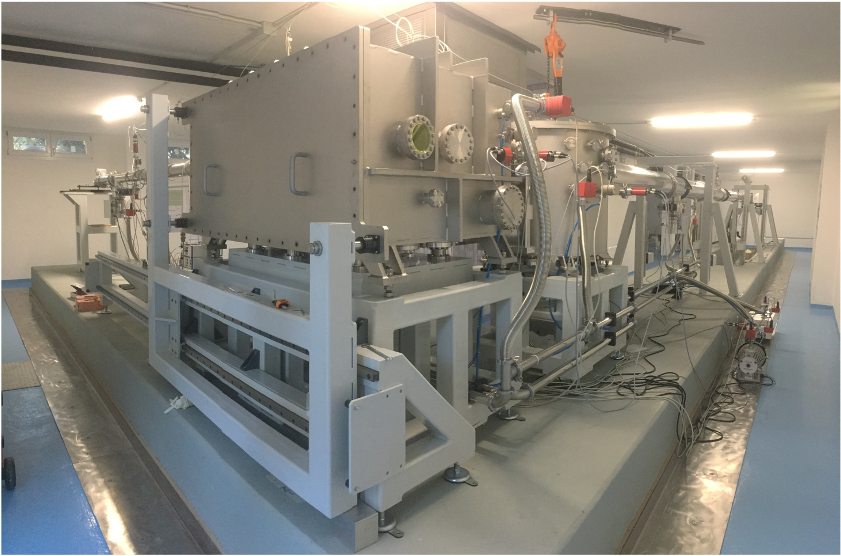}
   \caption{BEaTriX facility realized at the INAF-OAB laboratory. The vacuum chamber holding the optics component is visible in the foreground. Beam generation and handling occurs in the "short arm" on the left side. The 12~m tube (the "long arm") for X-ray propagation to the focal plane is on the right.}
    \label{fig:Beatrix_view}
 \end{figure}

The main goal of BEaTriX is to perform the acceptance tests of mirror modules (MMs) for the Advanced Telescope for High-ENergy Astrophysics (ATHENA) at their production rate of 2 MM/day, directly measuring their point spread function (PSF) and effective area (EA) in X-rays. ATHENA is the second large-class mission selected by the European Space Agency (ESA) within the Cosmic Vision Program, with a launch foreseen in the early 2030s (Nandra et al. \cite{Nandra2013}). The optics will consist of a large aperture X-ray mirror with a diameter of about 2.5~m, an EA of 1.4~m$^2$ at 1~keV, and a half-energy width (HEW: twice the median value of the PSF) of 5~arcsec at 1~keV (Bavdaz et al. \cite{Bavdaz2021}). The aperture of such a large X-ray telescope will be populated by 600 MMs, produced with the silicon pore optics (SPO) technology developed by ESA and cosine (Collon et al.~\cite{Collon2021}).
 
BEaTriX (Fig.~\ref{fig:Beatrix_view}), now operational at the Istituto Nazionale di Astrofisica (INAF), Brera Astronomical Observatory (OAB) in Merate (Italy), will soon start the systematic characterization of MMs for ATHENA. BEaTriX is also part of the Integrated Activities for the High Energy Astrophysics Domain (AHEAD) focused on X-ray optics (Burwitz et al.~\cite{Burwitz2018}). This paper reports the characterization of the X-ray parallel beam of BEaTriX and the first focused image obtained with an SPO MM, which was specifically provided by cosine to test the expanded beam properties.

\begin{figure}
   \centering
   \begin{tabular}{c}
   	\includegraphics[width= 0.48\textwidth]{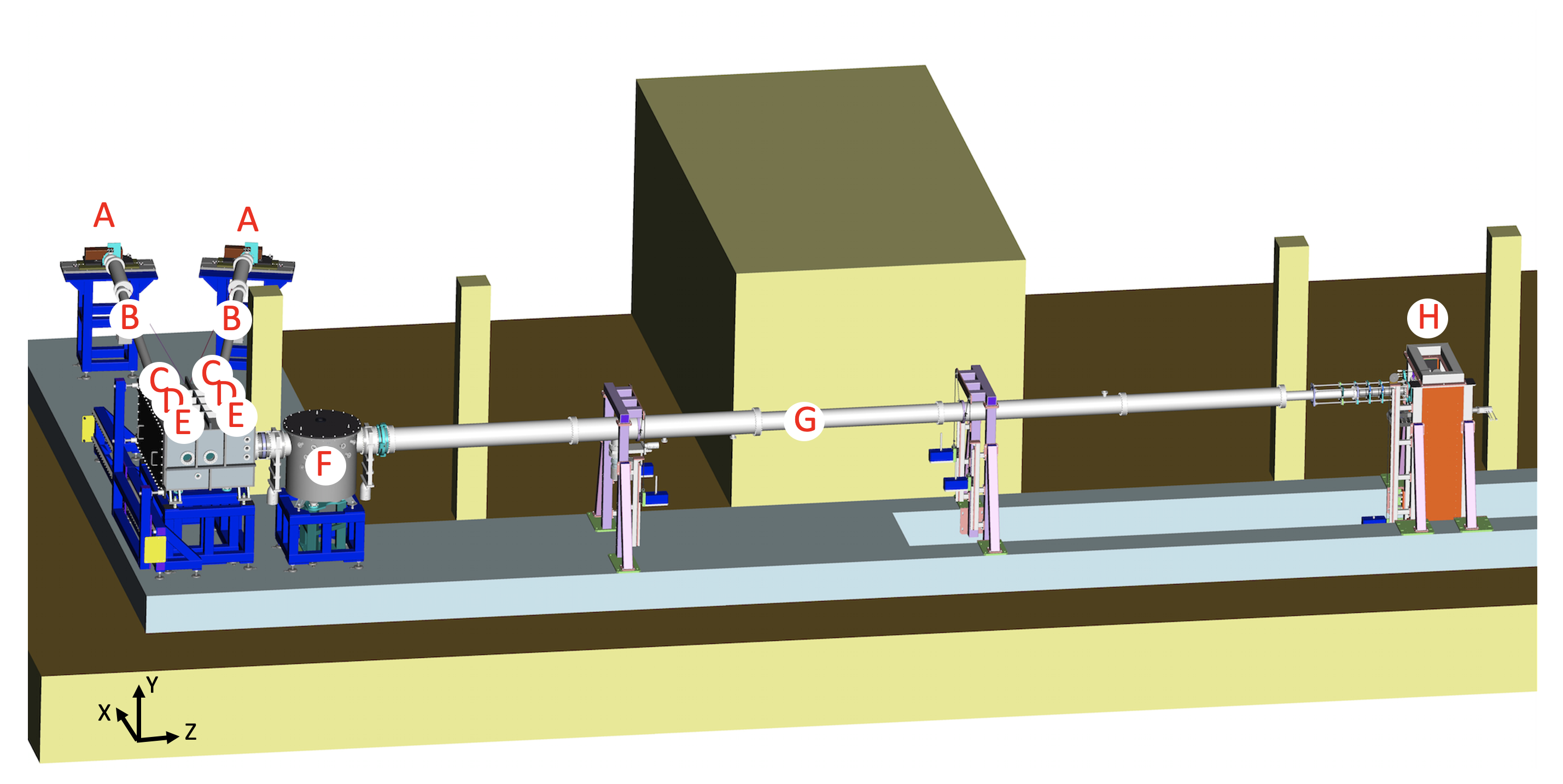}\\	
   	\includegraphics[width= 0.48\textwidth]{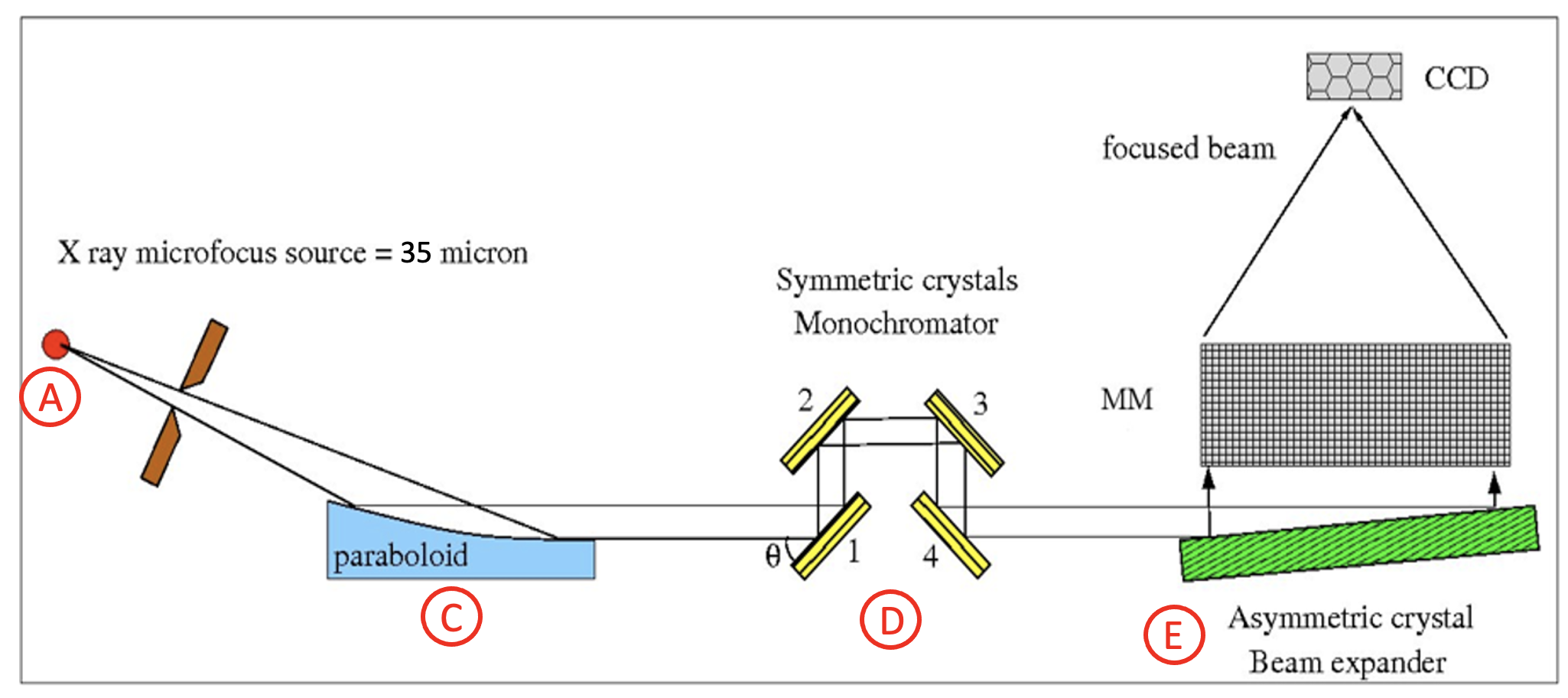}
   \end{tabular}
   \caption{Schematic of the BEaTriX facility: two-beamline layout and coordinate system, (top) and optical design for the 4.51 keV beamline (bottom). The optical components enable collimation, spectral filtering, and expansion of an X-ray beam.}
   \label{fig:Beatrix_layout}
 \end{figure}

\section{Facility description and X-ray beam handling}
\label{sect:descr}
\begin{figure}
   \centering
   \includegraphics[width= 0.48\textwidth]{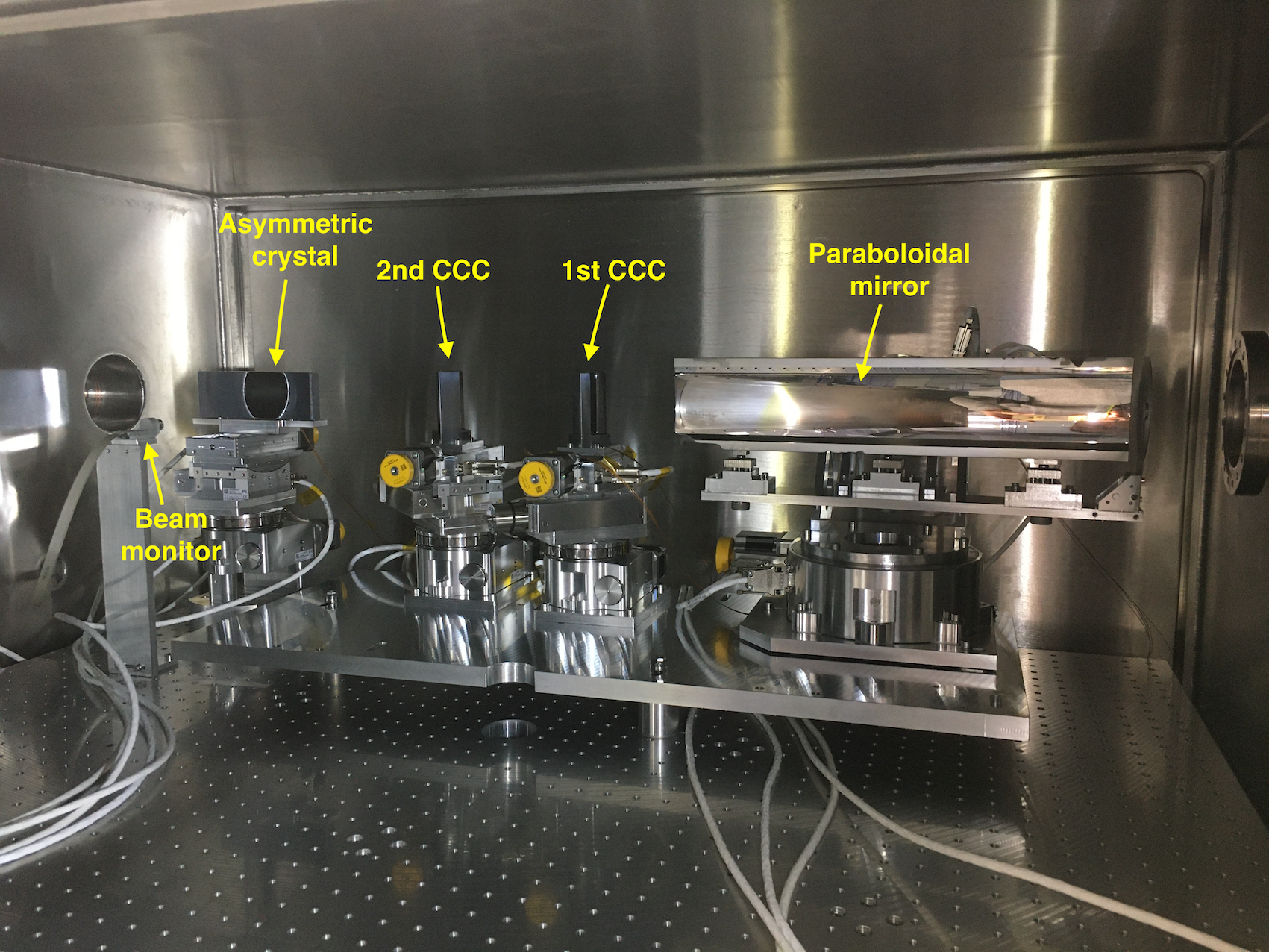}
   \caption{Optical components for beam handling, positioned on their motorization stages in the optical chamber.}
    \label{fig:vac_mot}
 \end{figure}

Tests of SPO MMs in BEaTriX are foreseen at the monochromatic energies of 4.51~keV and 1.49~keV. The 4.51~keV beamline of BEaTriX is currently operated and characterized. Its working principle is depicted in Fig.~\ref{fig:Beatrix_layout}; (A) X-rays are emitted by a micro-focus (35~$\mu$m full-width half maximum) source with a titanium anode set at 30 kV and 200 mA, (B) propagate through a vacuum tube, and (C) become parallel via reflection onto a paraboloidal mirror, accurately figured and polished in INAF-OAB to a 3~arcsec HEW level, then coated with a 30~nm platinum layer at the Denmark Technical University (DTU) to enhance its reflectivity. The mirror PSF was tested in X-rays at PANTER, before and after coating (Spiga et al. \cite{Spiga2021}; Vecchi et al. \cite{Vecchi2021}). The fluorescence line of titanium at 4.51~keV is subsequently filtered via a 4-fold diffraction monochromator based on silicon crystals (D), cut parallel to the (220) planes, and finally diffracted at about 90~deg off-surface by another silicon crystal, asymmetrically cut with respect to the (220) planes (E). The asymmetric diffraction ensures an approximately 50-fold horizontal expansion, making the final beam the same size as the asymmetric crystal (170 mm $\times$ 60 mm). This concept was previously tested for the calibrations of the Soviet Danish R\"ontgen Telescope (Christensten~et~al.~\cite{Chris1994}). In the 1.49~keV beamline, the asymmetric silicon crystal will be replaced by an ammonium dihydrogen phosphate crystal (ADP; Ferrari et al.~\cite{Ferrari2019}). Due to the intrinsic dispersive power of asymmetrically cut crystals (Sanchez Del Rio \& Cerrina \cite{SRio1992}), a very narrow energy band (0.05 $\div$ 0.1~eV) within the Ti-K$\alpha_1$ line has to be selected. To this end, the monochromator is based on two channel-cut crystals (CCCs) that can be used in two different configurations. If both CCCs are aligned at the diffraction peaks, the flux is maximized with some degradation in the horizontal collimation ("high-intensity" setup). In contrast, if the second CCC is rotated by 10~arcsec, so as to detune the diffraction peaks and make the system more selective in energy, then the total divergence is minimized at the expense of the flux intensity ("high-collimation" setup). Higher diffraction orders, such as (440), are prevented from reaching the monochromator by the cutoff of the collimating mirror at 6~keV (Spiga et al.~\cite{Spiga2016}). After expansion, a small beam monitor continuously records the intensity of the beam over time; the monitor is placed in a beam corner to minimize the obstruction. The optical components that enable beam filtering and expansion are displayed in Fig.~\ref{fig:vac_mot}.

\begin{figure}
   \centering
   \includegraphics[width= 0.48\textwidth]{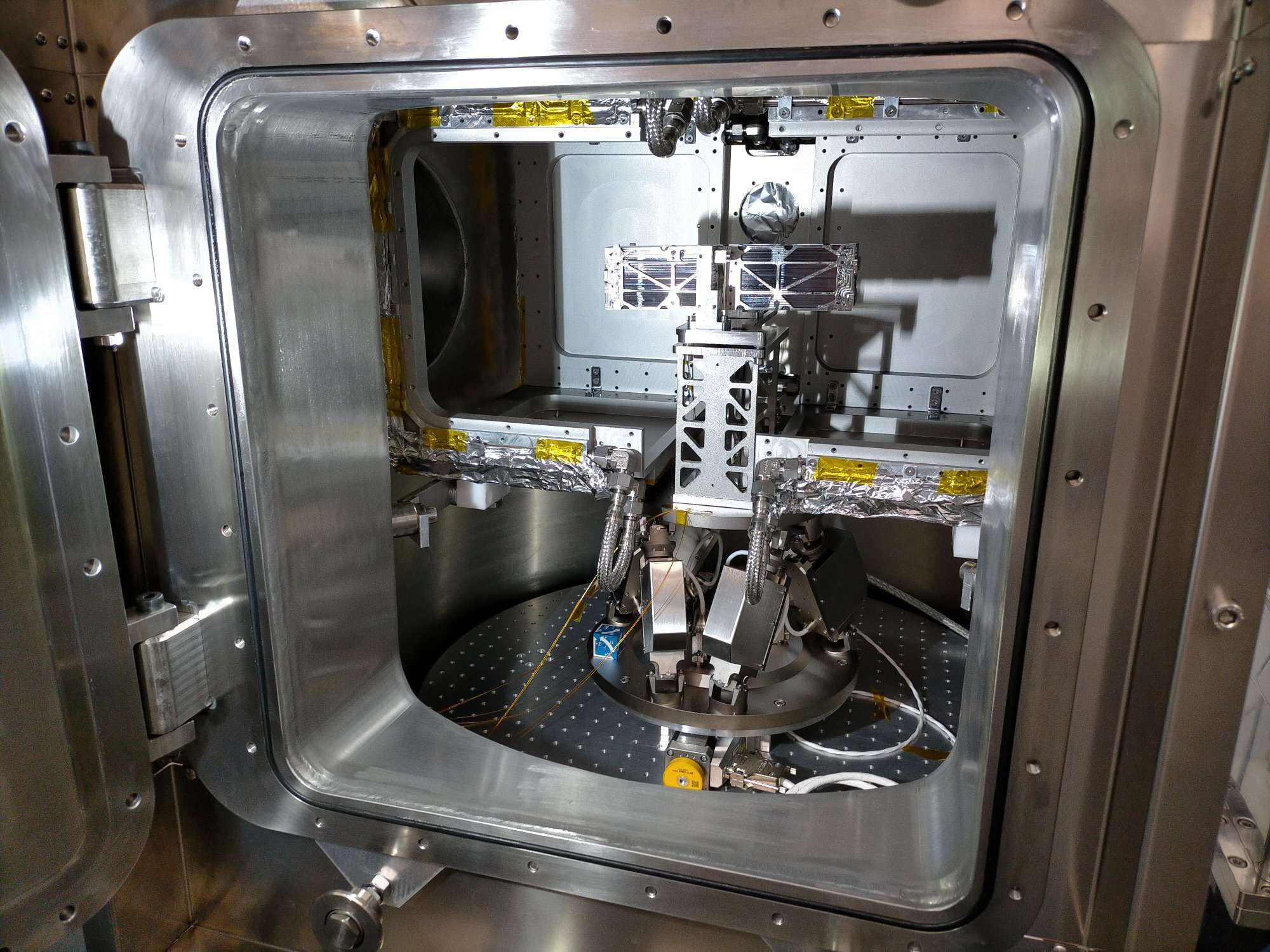}
   \caption{SPO MM (MM-0042) installed inside the vacuum chamber on the support, which is also used for the Hartmann plate. The support is mounted on the hexapod for alignment in the expanded beam.}
    \label{fig:exper_chamb}
 \end{figure}

The parallel and expanded beam then enters the experimental chamber (F), where it fully illuminates the aperture of the MM being tested (Fig.~\ref{fig:exper_chamb}); the MM is mounted onto a hexapod that allows us to align it in six degrees of freedom. The incidence plane, which corresponds to the radial plane of the MM, is set vertically to exploit the better collimation of the beam in the vertical direction, which is independent of the beam monochromation. Therefore, the beam is deflected downward and gets focused on a CCD camera placed at a 12 m distance (H). The CCD, equipped with a 27.6~mm square sensor with 13.5~$\mu$m pixels, is connected to the experimental chamber by a vacuum tube (G). A system of turbopumps keeps the full X-ray path in a vacuum at 10$^{-6}$ mbar; the whole vacuum system is partitioned by gate valves to evacuate and vent different sections separately. Notably, the experimental vacuum chamber can be isolated from nearby ones, so as to minimize the time needed for MM replacement. The CCD can be moved along the focus by 500~mm, laterally by 150~mm and vertically by 1400~mm in order to follow the beam deviation of modules with different incidence angles. The detector tower can be moved to abridge the CCD distance from the sample to 8 m or 10 m. Finally, the temperature of the MM being tested can be varied in the range -15$^{\circ}$C to +60$^{\circ}$C by means of a thermal box surrounding the optic inside the experimental chamber (Basso~et~al.~\cite{Basso2019}).
 
The components that handle the X-ray beam (C-D-E) are placed over precision motors for high vacuum (Fig.~\ref{fig:vac_mot}). For alignment purposes, the parabolic mirror can be rotated around the vertical axis and around the mirror normal; the CCCs can be rotated around the vertical axis or around the incident beam. The parabolic mirror was firstly aligned to its nominal position using a 3D coordinate metrology system. A subsequent alignment in the BEaTriX vacuum chamber was performed using a laser tracker. The alignment was refined with a Hartmann plate (Sect.~\ref{sect:characterization}) at the parabolic mirror exit. Each CCC was aligned at the peak of the beam intensity at its exit. The asymmetric crystal was also aligned by finding the incidence angle that maximizes the intensity of the expanded beam.

\section{X-ray beam characterization}
\label{sect:characterization}
We used a Hartmann test to measure the beam divergence (Idir et al.~\cite{Hartmann}). The Hartmann plate, an array of 400-micron-wide square holes, was made in stainless steel with a thickness of 250 $\mu$m. The center-to-center hole spacing is 2~mm vertical and 4~mm horizontal. The different spacing reflects the expectedly higher divergence in the horizontal direction: a larger separation between holes was left to avoid a possible superposition of beamlets. The plate was mounted on the MM holder in the experimental chamber and is illuminated by the expanded beam. At a 12~m distance, the CCD records the displacements of the beamlets from their nominal positions due to the residual wavefront distortions. The signal was integrated by the CCD for 30~minutes. Due to the beam being much wider than the CCD area, we took exposures at 21 positions, which were stitched together into a mosaic in a post-processing phase.
 
\begin{figure*}
   \centering
   \includegraphics[width= 0.9\textwidth]{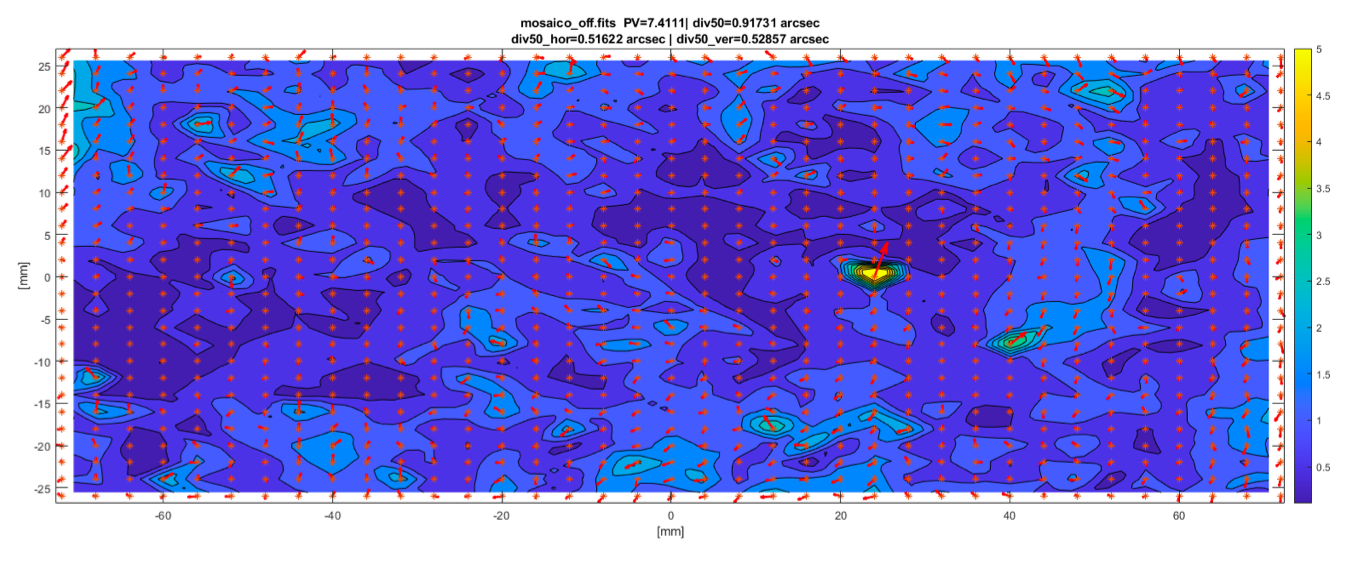}
   \caption{Divergence computation on the composition of 3x7 images. Colors represent the amplitude of the divergence in arcseconds and the red arrows the related direction. The "Div50" parameter denotes the median value of absolute angular deviation and it equals HEW/2. The bright spot was due to a cosmic ray that altered the count during the exposure.}
    \label{fig:centroids}
 \end{figure*}

\begin{figure*}
   \centering
   \includegraphics[width= 0.85\textwidth]{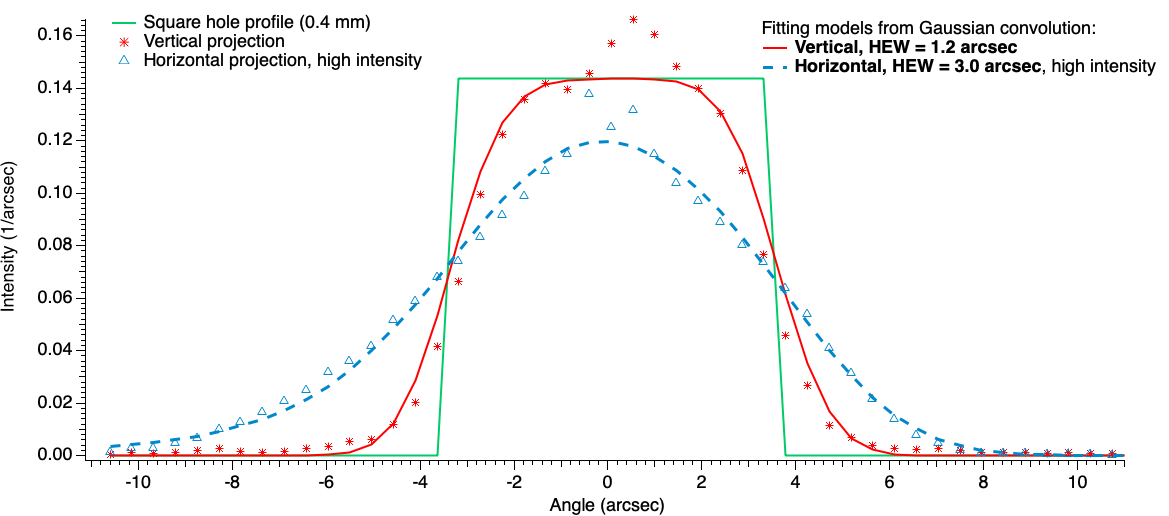}
   \caption{Intensity profiles of the beamlets out of the holes of the Hartmann plate, obtained in the high-intensity setup with the optimized pitch position of the parabolic mirror. The fitting models are convolutions of a squared hole profile to Gaussians that approximate the intrinsic beam PSF. All the curves are normalized to 1, and the vertical scale is in units of normalized power per arcsecond. Deviations from the Gaussian model near the top are due to the non-Gaussian distribution of the slope errors in the collimating mirror and the non-Gaussian rocking curve of the CCCs.}
    \label{fig:profiles}
 \end{figure*}

From the composite image, two components of the divergence HEW along the horizontal and the vertical direction can be calculated. First, measuring the displacements of the beamlet centroids from their nominal position (Fig.~\ref{fig:centroids}), we obtain the local slopes of the wavefront. Taking twice the median values of the horizontal and of the vertical slopes, we respectively obtain $\mathrm{HEW_{hor-centr}}$ and $\mathrm{HEW_{vert-centr}}$. These terms come from wavefront distortions due to residual misalignments of the optical components and low-frequency errors of the parabolic mirror. Second, fitting the intensity profiles of the beamlets along the two axes with an appropriate convolution model (Fig.~\ref{fig:profiles}), we derived $\mathrm{HEW_{hor-prof}}$ and $\mathrm{HEW_{vert-prof}}$. These two contributions arise from the imperfect spatial coherence of the beam due to (a) the nonzero, albeit small, size of the X-ray source, (b) the residual micro-roughness of the paraboloid, and (c) the bandwidth of the energies out of the monochromator. While $\mathrm{HEW_{vert-prof}}$ is affected only by the X-source dimension, all the factors above contribute to broaden $\mathrm{HEW_{hor-prof}}$. Therefore, the horizontal profile is typically broader than the vertical one.

 
To characterize the collimation of the expanded beam, we can add in quadrature the two contributions, which are assumed to be independent:
\begin{eqnarray}
	\mathrm{HEW_{hor}} & = & \sqrt{\mathrm{HEW^2_{hor-centr}+HEW^2_{hor-prof}}}\label{eq:HEWhor}\\
	\mathrm{HEW_{vert}} & = & \sqrt{\mathrm{HEW^2_{vert-centr}+HEW^2_{vert-prof}}}.\label{eq:HEWvert}
\end{eqnarray}
In the high-intensity setup (Fig.~\ref{fig:profiles}), with an expanded beam density of 50~ph/s/cm$^2$, the Hartmann test returned as total divergence:
\begin{eqnarray}
	\mathrm{HEW_{hor-beam}} & = & \left(3.45 \pm 0.25\right)~\mathrm{arcsec}\\
	\mathrm{HEW_{ver-beam}} & = & \left(1.65 \pm 0.25\right)~\mathrm{arcsec}.
\end{eqnarray}

  \begin{figure*}
   \centering
   \begin{tabular}{cc}
  	 \includegraphics[width= 0.4\textwidth]{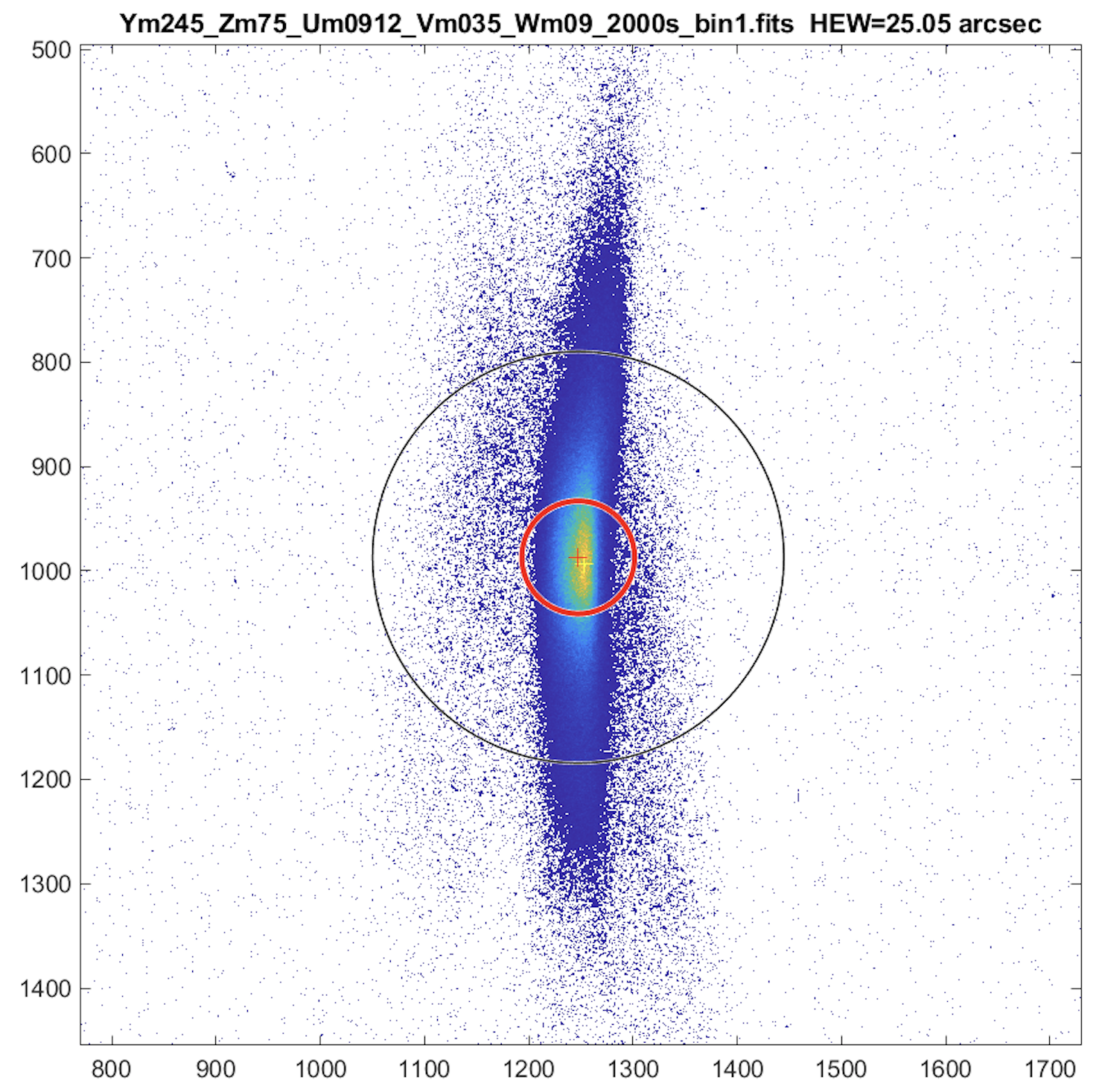}&
   	\includegraphics[width= 0.52\textwidth]{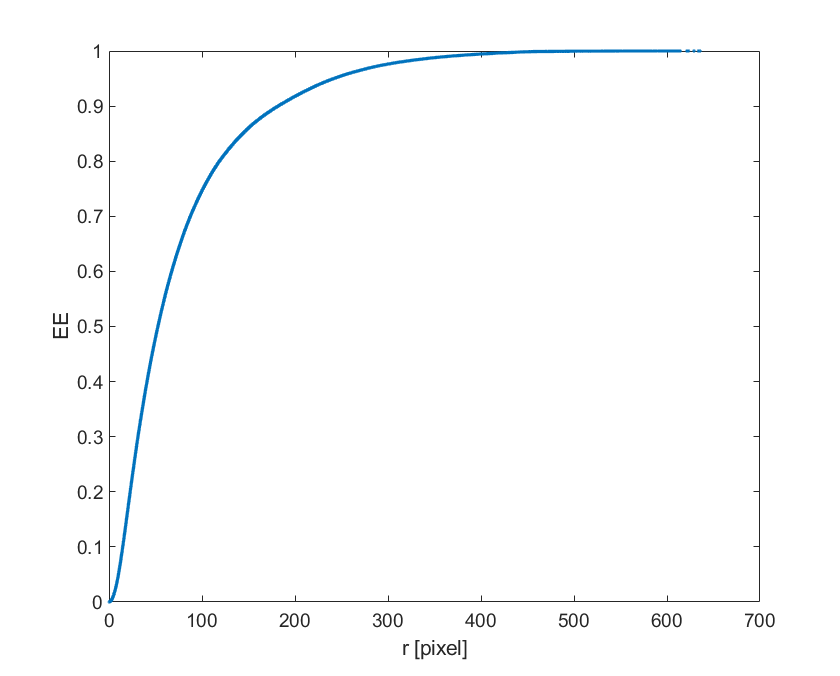}
   \end{tabular}
   \caption{Acquired image of MM-0042. Left: Pixel size is 13.5~$\mu$m, and the two circles correspond to the HEW and the diameter enclosing 90\% of the collected photons (W90). Right: Measured encircled energy (EE; the radial integration of the PSF) function of the MM-0042, used to retrieve the HEW and the W90 of the module.}
    \label{fig:MMmeasurement}
 \end{figure*}

The horizontal collimation can be improved in the high-collimation setup by rotating the second CCC by about 10~arcsec from the maximum intensity position. In this case, however, a mosaic image collection is very time-consuming due to the reduced flux. Therefore, we simply sampled the expanded beam at just one CCD position near the beam center. As a result, the $\mathrm{HEW_{hor-prof}}$ term is improved to just 2.3~arcsec, to be compared with a 2.2~arcsec value expected from simulations. On the other hand, the intensity was reduced to 10~ph/s/cm$^2$, which also agrees with the theoretical expectation and enables, in a 30-minute integration time, a fully meaningful measurement of the MM PSF and EA. 

From the Hartmann test in the high-collimation setup, we can derive the total HEW of the beam using the quadratic sum of the two terms (Eqs.~(\ref{eq:HEWhor}) and~(\ref{eq:HEWvert})):
\begin{eqnarray}
	\mathrm{HEW_{hor-beam}} & = & \left(2.7 \pm 0.25\right)~\mathrm{arcsec}\\
	\mathrm{HEW_{ver-beam}} & = & \left(1.65 \pm 0.25\right)~\mathrm{arcsec}.
\end{eqnarray}
During BEaTriX operation, either the high-intensity or the high-collimation configuration will be adopted, depending on the measurement accuracy required from each SPO MM being tested.

\section{X-ray measurements of a mirror module}
\label{sect:measurement}
An uncoated inner MM (named MM-0042) provided by cosine and tested at the X-ray Parallel Beamline Facility 2.0 at BESSY (Handick et al. \cite{Handlick2020}), was used as a validation test for the BEaTriX expanded beam. It consists of two identical X-ray optical units (XOUs), made from bare Si/SiO$_2$ plates, with an on-axis incidence angle of about $\theta_{\mathrm{inc}}$ = 0.3~deg, sufficiently small to reflect at 4.5~keV without any coating. Only the outer XOU was suitable for the measurement. Although this MM is not representative of the current production process of SPO optics, it is, at the present time, the only MM available for tests at 4.5~keV. Due to the grazing incidence, figure imperfections are expected to mostly cause PSF elongation in the incidence plane, which is set vertical in the BEaTriX setup. The vertical direction, being the one with better collimation, is also the one with higher sensitivity to the PSF of the MM.

The MM was mounted in the vacuum chamber (Fig.~\ref{fig:exper_chamb}) and pre-aligned mechanically; then the CCD was placed at the position expected from a 4$\theta_{\mathrm{inc}}$ deviation of X-rays downward. The alignment in pitch and yaw was optimized by maximizing the reflected X-ray flux; the maximum corresponds to the on-axis illumination because the obstructions by the ribs and the membranes are minimized in an SPO MM. To find the best focus distance, a scan along the beam direction was carried out finding the minimum of the image width in the horizontal direction because, due to the relevant vertical elongation of the PSF, the HEW minimum is not a good metric for the best focus. The focused image was integrated for a total time of 2000~s.

The focused image of the MM (Fig.~\ref{fig:MMmeasurement}) returned a HEW value of (25.24 $\pm$ 0.89)~arcsec, in agreement with the value measured at BESSY. The computed EA was (6.84 $\pm$ 0.35)~cm$^2$, evaluated with a collection of at least 10$^5$ photons. The value is to be compared with a theoretical value of 6.72 cm$^2$. This validates the proper functioning, alignment, and calibration of BEaTriX.

\section{Conclusions}
\label{sect:conclusions}
The first beamline of the BEaTriX facility, operated with a flux of 50 photons/s/cm$^2$ at 4.5~keV, is now ready to test ATHENA MMs, an activity that will be given the highest priority in the next few years. In the meantime, the second line at 1.49 keV will be implemented. BEaTriX was built to test X-ray optics with a parallel beam of 170 mm $\times$ 60 mm in size, with a focal length in the range 7.8 $\div$12.2~m.

The first light of the direct beam showed collimation and intensity in line with the expectations. Measurements show that the beam directionality is stable on timescales of hours, largely within the typical time foreseen for measurements (30 minutes). The recalibration and alignment of the components on longer timescales (e.g., days) will be assessed in the forthcoming months, increasing the statistics of the recorded data.

The performance of the beamline was tested with an early prototype of inner radius SPO MM, which confirmed that both the angular resolution and the EA can be reliably measured. The optical quality of this MM did not fully exploit the BEaTriX capabilities, which can characterize optics with a quality ten times better, but it demonstrates the great potential of the facility to test the ATHENA MMs (focal length, angular resolution, and EA) at the requested rate of 2 MMs/day.

\begin{acknowledgements}
	The project is financed by ESA (contract \#4000123152/18/NL/BW), AHEAD (grants \#654215 and \#871158), ASI (grant \#2019-27-HH.0), and INAF internal funds. 
\end{acknowledgements}

%
%


\end{document}